\documentclass[%
reprint,
showpacs,
footinbib,
 amsmath,amssymb,
 aps,
]{revtex4-1}

\usepackage{graphicx}
\usepackage{dcolumn}
\usepackage{bm}
\usepackage{siunitx}
\usepackage{amsmath}

\begin{document}

\preprint{APS/123-QED}
\title{Frequency and amplitude dependent population dynamics during cycles of feast and famine}

\author{Jason Merritt$^{1,2}$}
\author{Seppe Kuehn$^{1,2,3}$}%
 \email{To whom correspondence should be addressed: seppe@illinois.edu}
\affiliation{%
$^1$Center for the Physics of Living Cells, University of Illinois at Urbana-Champaign, Urbana, IL 61801, USA. \\$^2$Department of Physics, University of Illinois at Urbana-Champaign, Urbana, IL 61801, USA. \\$^3$Center for Biophysics and Quantitative Biology, University of Illinois at Urbana-Champaign, Urbana, IL 61801, USA. 
}%

\date{\today}

\begin{abstract}
In nature microbial populations are subject to fluctuating nutrient levels.  Nutrient fluctuations are important for evolutionary and ecological dynamics in microbial communities since they impact growth rates, population sizes and biofilm formation.  Here we use automated continuous-culture devices and high-throughput imaging to show that when populations of \textit{Escherichia coli} are subjected to cycles of nutrient excess (feasts) and scarcity (famine) their abundance dynamics during famines depend on the frequency and amplitude of feasts.  We show that frequency and amplitude dependent dynamics in planktonic populations arise from nutrient and history dependent rates of aggregation and dispersal.  A phenomenological model recapitulates our experimental observations.  Our results show that the statistical properties of environmental fluctuations have substantial impacts on spatial structure in bacterial populations driving large changes in abundance dynamics.
\end{abstract}

\pacs{87.23.Cc, 87.10.Vg, 87.18.-h, 87.18.Vf, 87.18.Fx, 87.18.Ed}
\maketitle

In nature, microbial populations are subjected to temporally and spatially variable environments.  In ecosystems including oceans, lakes, and soils, limiting nutrients are present as patches or particles \cite{Simon:2002vz,Datta:2016bj} and at low concentrations \cite{Dittmar:2014gd}.  As a result, nutrient conditions are believed to be dynamic with microbes experiencing periods of nutrient excess and starvation on multiple timescales \cite{Seymour:2017bd,Savageau:1983uo}.  

In many contexts bacterial populations also transition between free-floating aggregates \cite{Schleheck:2009jr} or surface-attached biofilms \cite{McDougald:2011bz} and dispersed planktonic populations \cite{Laganenka:2016dp}. Nutrient conditions affect the development of this spatial structure.  For example, increases in nutrient availability drive biofilm dispersal in some species \cite{Sauer:2004dja,Schleheck:2009jr}, and bacterial populations resident in biofilms enter stationary phase \cite{Ito:2008fz} while becoming more resistant to antibiotics \cite{Ito:2009jr}.  However, our understanding of how the statistics of environmental fluctuations interact with the formation and dispersal of spatial structure in microbial populations is limited.  

In this Letter we present quantitative measurements of the population dynamics of \textit{Escherichia coli} cycling between conditions of nutrient excess (feasts) and starvation (famine).  We vary the frequency and amplitude of nutrient fluctuations and observe a strong dependence of the abundance dynamics on both variables.  We find that nutrient fluctuations with higher frequency and amplitude drive faster abundance dynamics in planktonic populations.  Further, populations subjected to nutrient fluctuations on timescales shorter than \num{2} days exhibit memory on a timescale that exceeds a generation time.  Our data, in combination with a simple model of community dynamics, show that these phenomena arise from a history and substrate dependence in the dispersal of aggregated or adherent bacterial populations.  Finally, we document a concomitant frequency and amplitude dependence in the lag-phase duration of bacterial populations.

\begin{figure*}
\includegraphics[width=\textwidth]{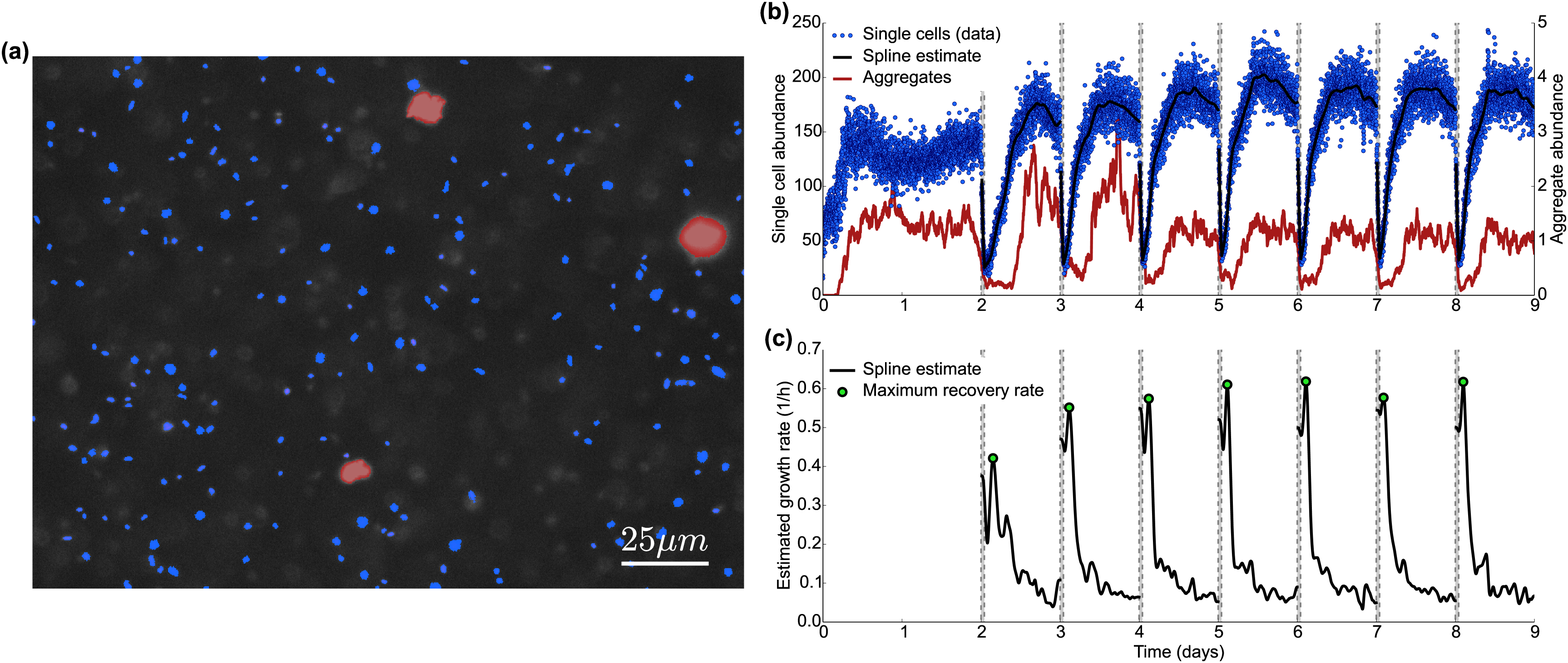}
\caption{(Color online)  \textit{E. coli} abundance dynamics in fluctuating nutrient conditions. (a) Example epi-fluorescence image showing single cells (blue) and aggregates (red) detected by image processing. (b) Number of planktonic cells (blue points) and cell aggregates (red line) detected per image by automated measurement, with aggregate abundances smoothed by a \num{1} hour rolling average.  Dashed vertical lines indicate regions of time where a washout event occurred (\num{1} hour duration).  Black lines indicate spline estimates of planktonic population abundances. (c) Instantaneous growth rate for planktonic population estimated from the spline fits shown in (b).  Green dots indicate maximum growth rate during recovery.   \label{fig1}}
\end{figure*}

We use custom continuous-culture devices coupled to epi-fluorescence microscopes which image fluorescently labeled \textit{E. coli} at the single-cell level [Fig. \ref{fig1}(a)].  Our continuous-culture devices permit long-term automated imaging to measure population dynamics on timescales of minutes for periods of weeks \cite{Merritt:2016ea}. We maintain a \SI{20}{\milli\liter} culture of bacteria in chemostat conditions while a pump draws samples from the culture once per minute and passes them through a micron-scale glass capillary where imaging occurs.  We use a strain of \textit{E. coli} expressing \textit{dTomato} constitutively from the chromosome. Populations are grown in M\num{63} minimal medium at \SI{30}{\celsius} with low levels of carbon (\SI{0.04}{\percent} w/v, \num{2.2} mM glucose).  Prior to an experiment, populations are initiated from single colonies and grown in a batch culture and then transferred to the continuous-culture devices operating at a basal dilution rate of $D=$ \SI{0.08}{\per\hour} (doubling time $\tau_d =$ \SI{8.66}{\hour}) for \num{48} hours to acclimate to the conditions of slow but continuous growth.  We operate six chemostats in parallel.

Following the acclimation period, the continuous-culture devices alternate between long, fixed periods of chemostat operation at the basal dilution rate (famine) and short ``washout events" where \num{90}-\SI{99}{\percent} of the population is replaced with fresh medium over the course of one to two hours (feasts) [Fig. \ref{fig1}(b)]. Washout events simultaneously reduce the population by a factor of \numrange{10}{100}, depending on the amplitude, and increase the substrate (glucose) concentration from a few micromolar \cite{Wick:2002bd} to approximately \num{2} mM, resulting in periods of rapid growth as the population recovers to its steady state abundance [Fig. \ref{fig1}(b)].  

During cycles of feast and famine we perform automated imaging once per minute on samples drawn from the growing bacterial population.  During periods of famine we observe both planktonic (single-cell) populations and aggregated cells [Fig. \ref{fig1}].  From the size of the aggregates (Supplemental Material Fig. S11 \footnote{See Supplemental Material in ancillary files, which includes Refs. \cite{Merritt:2016ea,LevinReisman:2010jp,Scott:2010cxa,Senn:1994jf,Ito:2008fz}, for details on experimental protocol, model parameters, image segmentation, growth rate, cell size, and cell aggregation.}) we estimate that, at steady state, the numbers of planktonic and aggregated cells are of the same order (\num{1e8} mL$^{-1}$).  During washout events the planktonic population declines by \num{10}- to \num{100}-fold and the number of aggregates falls to nearly undetectable levels.  Subsequently, with the chemostat operating at the basal dilution rate, the planktonic population rapidly returns to its steady state value.  During this recovery we measure the instantaneous growth rate of the planktonic population.  We find this time dependent growth rate exhibits a peak early in the recovery [Fig. \ref{fig1}(c)].  We report this maximum recovery growth rate [green points in Fig. \ref{fig1}(c)].  The population of aggregates remains low ($<$\num{0.1} per image) until the planktonic population growth rate declines below \SI{0.2}{\per\hour} and then begins to recover \cite{Merritt:2016ea}.
   
\begin{figure}
\includegraphics[width=\columnwidth]{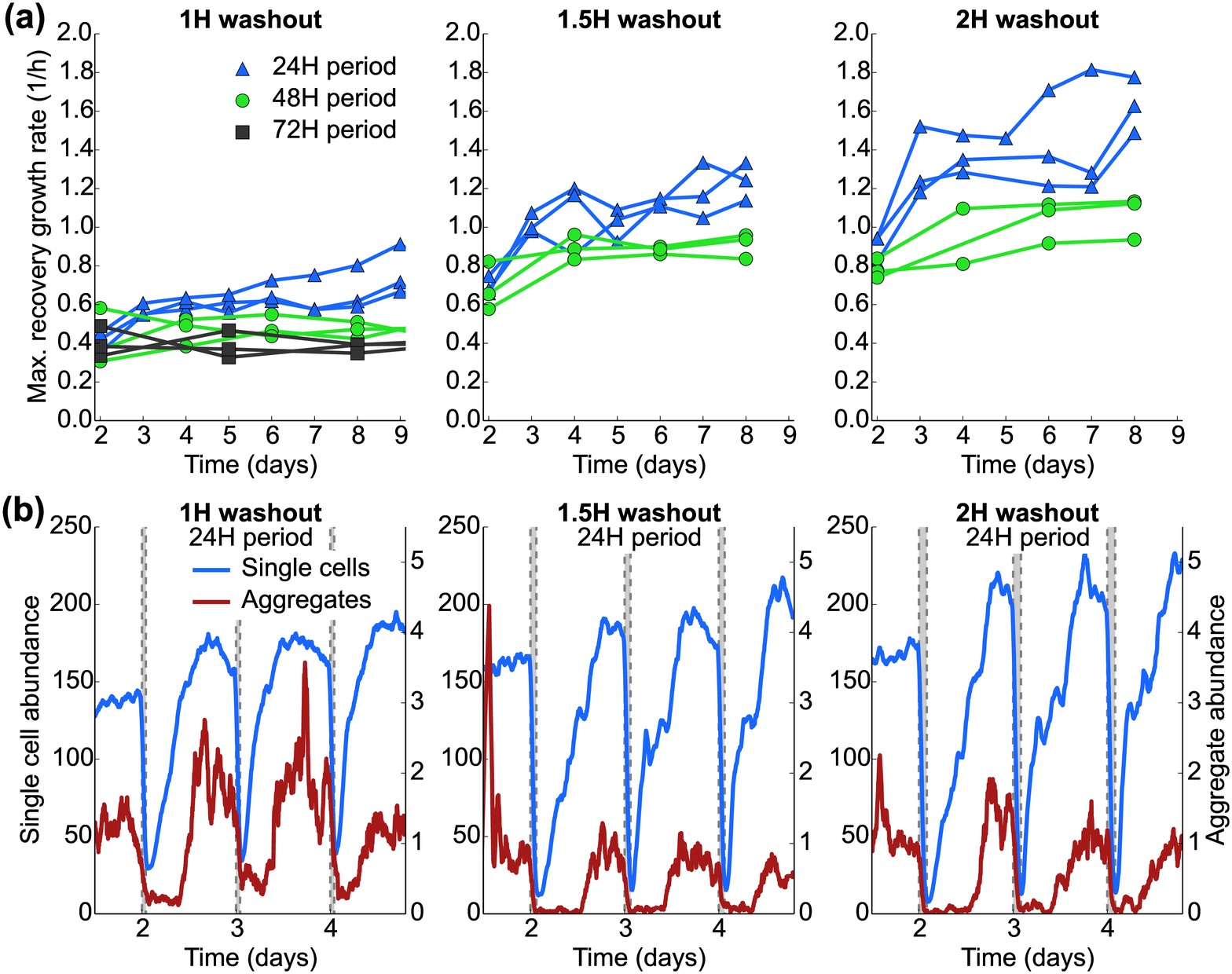}
\caption{(Color online) Frequency and amplitude dependent abundance dynamics.  (a) Maximum growth rates of planktonic populations observed during recovery from a washout event for washouts that occurred every \num{72} hours (black), \num{48} hours (green) and \num{24} hours (blue) with durations varying from \num{1} hour (\num{1/10} dilution, left panel), \num{1.5} hour (\num{1/30} dilution, middle panel) and \num{2} hour (\num{1/100} dilution, right panel).  For each condition three independent replicates are shown.  Legend in left panel applies to all panels in (a).  (b) Example per-image abundances of planktonic populations ($N$) and aggregates ($A$) for systems experiencing washouts every \num{24} hours with amplitudes of \num{1} (left), \num{1.5} (middle) and \num{2} (right) hours respectively. Each abundance time series is smoothed with a \num{1} hour rolling average.   \label{fig2}}
\end{figure}

To study the frequency dependence of the observed abundance dynamics we performed \num{1} hour washout events which reduced the population by \num{10}-fold with periods ranging from every \SI{72}{\hour} to every \SI{24}{\hour} from the start of one washout event to the next.  We find that the rate of recovery of the planktonic population following a washout event increases the more frequently washout events occur [Fig. \ref{fig2}(a)].  The change in recovery rate occurs rapidly (by the second washout event), so we conclude that the change in population dynamics is the result of phenotypic processes rather than genetic mutations sweeping through the population \cite{Merritt:2016ea}. 

We next performed a series of experiments where the amplitude of the washout event was varied.  Washout events of larger amplitude occur over a longer period of time, resulting in a larger fraction of the population being washed out and a modestly higher final substrate concentration ($\sim$\num{2.2}mM rather than $\sim$\num{2}mM).  We performed washout events with durations of \SI{1.5}{\hour} and \SI{2}{\hour} and periods of \SI{24}{\hour} and \SI{48}{\hour}.  We find that larger amplitude washout events result in substantially faster growth during the recovery [Fig. \ref{fig2}(a)], with maximum recovery rates as high as \SI{1.4}{\per\hour}.  This rate exceeds previously measured biomass growth rates for \textit{E. coli} in glucose minimal media by at least a factor of four \cite{Scott:2010cxa}, suggesting that our measured planktonic population growth rate cannot be the result of cell division alone.  Both the frequency and amplitude dependent dynamics observed via imaging were corroborated by concurrent optical density measurements \cite{Note1}.  Fig. \ref{fig2}(a) is the central finding of this Letter.

One possible explanation for slow growth rates in low frequency perturbation conditions is the presence of phenotypic heterogeneity in the population such as dormant or persistent cells increasing their relative abundance with increasing famine duration \cite{Patra:2013if}.  To test this hypothesis we sampled chemostat populations every \num{12} hours over a \num{60} hour period of famine and used a previously developed assay to detect persistent cells by measuring the time for colonies to appear on agar plates \cite{LevinReisman:2010jp}.  We found no evidence of persisters in our experiment at relative abundances greater than approximately \SI{1}{\percent} regardless of the famine duration. Instead, the time for colonies to form on agar plates was approximately normally distributed regardless of when we sampled the population from the chemostat.  However, we did observe a monotonic dependence of the average time to colony formation (lag time) with the duration of the famine, as well as a decrease in the time to colony formation with increasing washout amplitude (Supplemental Material Figs. S2 and S7 \cite{Note1}).  These results show that the average time for cells to resume growth after a famine decreases with both the frequency and amplitude of environmental perturbations. 

We next considered the role cell aggregation plays in the dynamics shown in Fig. \ref{fig2}.  We performed an experiment where the basal dilution rate between washout events was set to zero.  In this condition populations do not continually grow between washout events but enter stationary phase as they would in batch culture.  Previous measurements showed that in batch culture lag phase duration also increases with starvation duration \cite{LevinReisman:2010jp}.  However, the maximum rate of recovery from washout events for planktonic populations in this condition is uniformly slow (maximum recovery rates $\sim$\SI{0.3}{\per\hour}), with no frequency or amplitude dependence (Supplemental Material Fig. S15 \cite{Note1}).  Critically, we observe little or no aggregation in batch culture conditions, with the entire population being planktonic \cite{Note1,Merritt:2016ea}.  This result strongly suggests that the presence of aggregated cells is necessary for the high maximum recovery rates shown in Fig. \ref{fig2}.  Under this premise, fast recovery rates exhibited by planktonic populations would be driven by the dispersal of aggregated or potentially adherent cells in the community.  

In light of these results, we sought a model to describe the frequency and amplitude dependent abundance dynamics we observe in bacterial populations growing in fluctuating nutrient conditions which captured the formation and dispersal of aggregated populations.  Our model considers populations of planktonic cells $N(t)$ and cells in free floating aggregates or adhered to the vessel $A(t)$.  We assume planktonic cells grow at a rate determined by the instantaneous substrate concentration $S(t)$.  Aggregates have a characteristic size of approximately \num{100} cells which we determined from imaging \cite{Note1}.  Given the large difference in apparent growth rates for planktonic populations between \num{1} hour and \num{2} hour washout events we reasoned that the dispersal rate of $A$ should increase with higher levels of available substrate $S$, an assumption which is supported by the literature \cite{Sauer:2004dja} and our observation that the size of aggregates decreases after washout events (Supplemental Material Fig. S12 \cite{Note1}).  To capture the history dependent recovery rates we assume that the rate of dispersal also depends on the duration of the famine, with longer famines resulting in lower dispersal rates, possibly due to maturation \cite{Reisner:2003uo}.  Finally, we assume that the $A$ population consumes no substrate since bacteria in biofilms have been shown to be in stationary phase \cite{Ito:2008fz}.  From these assumptions we construct the following dynamical model:

\begin{multline}
\dot{N} = \mu(S)N - DN  -\alpha_1(1 - f(S))N \\
+\alpha_2\frac{Q}{1+Q}f(S)AY_{NA},
\label{Ndynamics}
\end{multline}  

\begin{equation}
\dot{A} = \alpha_1(1 - f(S))\frac{N}{Y_{NA}} - \alpha_2 \frac{Q}{1+Q} f(S) A - D_{eff}A,
\label{Adynamics}
\end{equation}  

\begin{equation}
\dot{S} = (S_r - S)D - \frac{\mu(S)}{y}N.
\label{Sdynamics}
\end{equation}  

Here $\mu(S) = \frac{\mu_mS}{K+S}$, $D$ is the dilution rate of the chemostat, and $\alpha_1$ is the rate of $A$ formation from planktonic cells, modulated by substrate levels via $f(S)$.  $\alpha_2$ is the rate of $A$ dispersal and is modulated by substrate levels and $Q$, a variable that describes maturation of $A$ by reducing dispersal as the duration of starvation increases.  $Q$ increases when nutrients are replete ($\dot{Q} = aQ$ for $S>S_c$) and decreases when nutrients are scarce ($\dot{Q} = aQ$ for $S<S_c$) in an autocatalytic fashion.  We chose autocatalytic dynamics for this variable because it is likely driven by a synthesized molecular species \cite{Laganenka:2016dp} or gene products \cite{Ito:2008fz} but the exact nature of the dynamics is not critical for the model.  $f(S)$ captures the increase in aggregate dispersal rate with substrate levels and is a monotonic increasing function of $S$ ($0 \leq f(S) \leq 1$, $f(S_r) = 1$) which we take to be linear above some threshold $S_{th}$ \cite{Note1}.  $Y_{NA}$ is the characteristic size of the aggregates ($\sim$\num{100} cells) and $y$ is the growth yield for \textit{E. coli} on glucose.  $D_{eff} = \rho D$, with $0 \leq \rho \leq 1$, provides a proxy for populations adhered to the vessel and therefore not removed by dilution.  

Our data permits us to constrain many of the model parameters, including the rate of aggregation ($\alpha_1$), and the rates of accumulation and degradation of $Q$ ($a$ and $b$), $\mu_m$, $K$, and $y$ have been measured previously \cite{Senn:1994jf}. We make analytical arguments to estimate the dispersal rate $\alpha_2$ \cite{Note1}.  The substrate concentrations $S_c$ and $S_{th}$ are not known, but our conclusions are not contingent on the specific values of these parameters, and all other parameters are under experimental control.  A full description of the model and detailed reasoning for the parameters used in our simulation is given in the Supplemental Material \cite{Note1}.  

\begin{figure}
\includegraphics[width=\columnwidth]{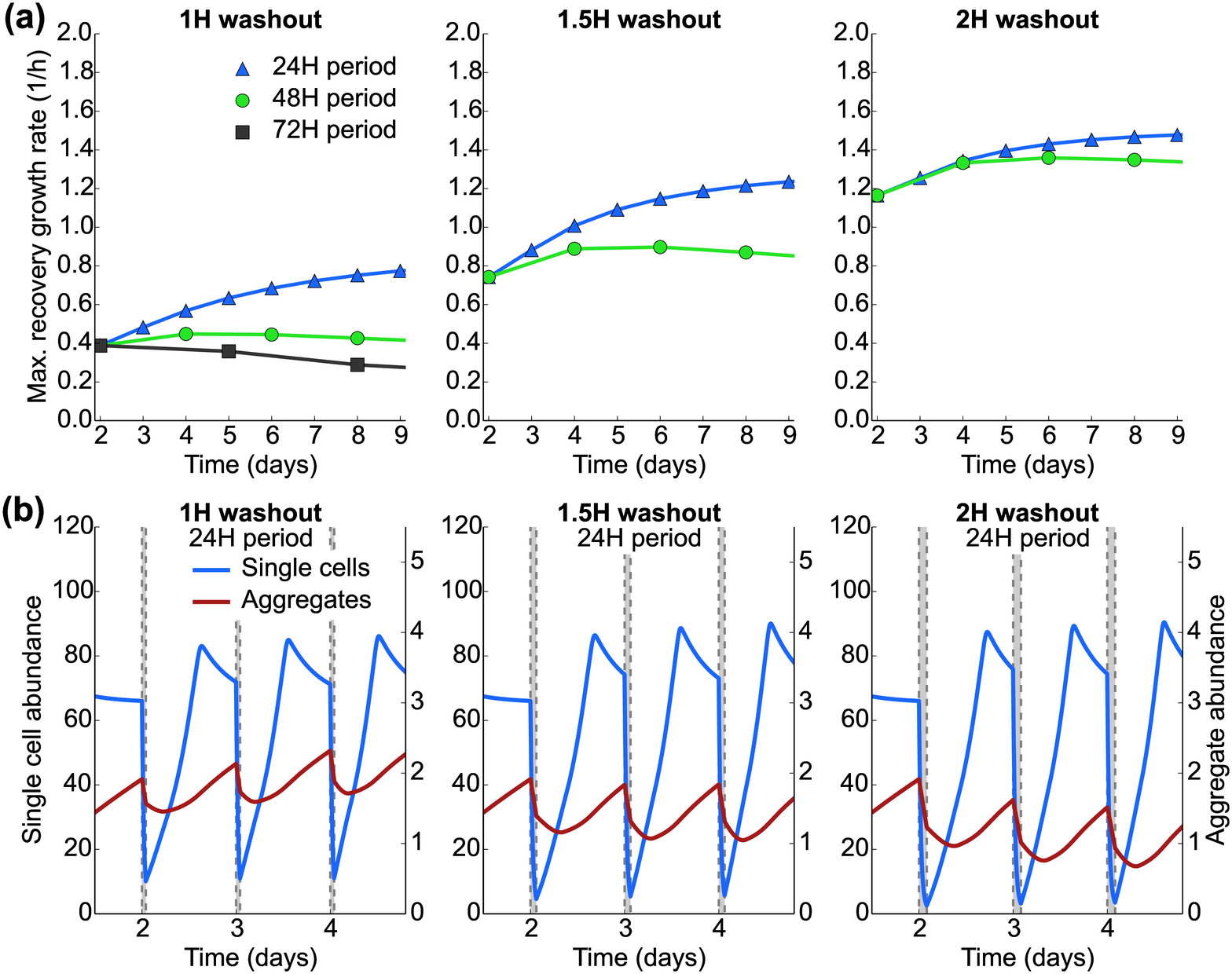}
\caption{(Color online) Simulated abundance dynamics.  Numerical integration of a model describing planktonic ($N$) and aggregated or adherent ($A$) population dynamics (see main text for details).  Panels are identical to Fig. \ref{fig2}. (a) Shows the maximum growth rate of planktonic populations computed during recovery from a washout event for washouts that occurred every \num{72} hours (black), \num{48} hours (green) and \num{24} hours (blue) with durations varying from \num{1} hour (\num{1/10} dilution, left panel), \num{1.5} hour (\num{1/30} dilution, middle panel) and \num{2} hour (\num{1/100} dilution, right panel).  (b) Simulated abundance dynamics of planktonic populations ($N$) and aggregates ($A$) for systems experiencing washouts every \num{24} hours with amplitudes of \num{1} (left), \num{1.5} (middle) and \num{2} (right) hours respectively.  \label{fig3}}
\end{figure}

We numerically integrated Equations (\ref{Ndynamics}), (\ref{Adynamics}) and (\ref{Sdynamics}) and computed the maximum recovery rate as a function of the frequency and amplitude of nutrient fluctuations.  We find that the model recapitulates the core features of our experimental observations, namely the frequency and amplitude dependence of the planktonic population abundance dynamics (Fig. \ref{fig3}).  The model shows that the dispersal of aggregated or adherent populations can drive the very high planktonic population growth rates we observe experimentally.  

We have shown that aggregation or adherent populations respond to increases in nutrient concentrations in a frequency and amplitude dependent fashion.  In contrast to recent studies of chemotaxis driven aggregation \cite{Laganenka:2016dp}, the dynamics we observe occur despite the fact that our strain lacks the \textit{flu} gene which encodes an adhesion factor (\textit{Ag43}) known to drive aggregation at \SI{37}{\celsius}.  We suspect that the adhesion dynamics are driven by curli-mediated cell-cell adhesion, which is known to occur at the lower temperature used in this study (\SI{30}{\celsius}) \cite{Olsen:1989iv,Laganenka:2016dp}.

It is increasingly clear that non-planktonic bacterial populations are central to metabolic \cite{Ito:2009jr}, evolutionary \cite{Shapiro:2012jp} and ecological processes \cite{Datta:2016bj} in a range of habitats.  Our study demonstrates that the statistical properties of environmental fluctuations have strong impacts on the lifestyle of bacterial populations which in turn drive rapid changes in abundance dynamics.  In the future, it will be important to investigate the eco-evolutionary origins of the frequency and amplitude dependent dynamics observed here.

\begin{acknowledgments}
This work was supported in part by the National Science Foundation Physics Frontiers Center Program (PHY 0822613 and PHY 1430124).
\end{acknowledgments}

\bibliography{Merritt2016ResilienceBib,JasonBib}

\end{document}